\documentclass[twocolumn,showpacs,prd,floatfix,axodraw]{revtex4}
\usepackage{mathrsfs}
\usepackage{graphicx,booktabs,bm}
\usepackage{subfigure}
\usepackage{overpic}
\usepackage{color}
\usepackage{amssymb}

\usepackage{mathrsfs,bm,amsmath,amssymb}
\usepackage{longtable,lscape}
\usepackage{txfonts}
\usepackage{amssymb}
\usepackage{indentfirst}
\usepackage{graphicx,,booktabs}
\usepackage{multirow}
\usepackage{color}
\usepackage{amssymb}

\definecolor{cover}{rgb}{0.77,0.87,0.88}
\definecolor{blueone}{rgb}{0.1,0.1,.7}
\definecolor{citec}{rgb}{0.14,0.47,0.09}
\definecolor{two}{rgb}{0.0,0.5,0.}
\definecolor{three}{rgb}{.5,.1,0.15}
\usepackage[bookmarks=true,bookmarksopen=false,plainpages=false,breaklinks=true,
bookmarksnumbered=true,hypertexnames=false,
filecolor=blue,urlcolor=blueone,menucolor=three,
linkcolor=red,citecolor=blue, colorlinks,
anchorcolor=blue,runcolor=pink,frenchlinks=red
pdfstartview=FitH,pdftitle=title,%
pdfauthor=author]{hyperref}

\begin{document}
	\title{Search for the $D^{*}\bar{D}^{*}$ Molecular State $X_{2}(4013)$ in $K^{-}p$ and $pp$ Collisions}

	\author{Min Yuan$^{1}$}
        \author{Bo Nan Zhang$^{2}$}
	\author{Yin Huang$^{1}$}\email{huangy2019@swjtu.edu.cn}

	\affiliation{$^{1}$School of Physical Science and Technology, Southwest Jiaotong University, Chengdu 610031,China}
    \affiliation{$^{2}$College of Physics and Electronic Information,
Inner Mongolia Normal University, Hohhot, 010022, Inner Mongolia, China}

	\date{\today}
	
\begin{abstract}
Motivated by the interpretation of $X(3872)$ as a $D\bar{D}^{*}$ molecular state, heavy-quark spin symmetry predicts a spin-2 partner, $X_{2}(4013)$, which can be regarded as a $D^{*}\bar{D}^{*}$ molecule with quantum numbers $J^{PC} = 2^{++}$. Its experimental confirmation, however, remains elusive. In this work, we investigate the production mechanisms of $X_{2}(4013)$ in the reactions $K^{-}p \to D_{s}^{-}\Lambda_{c}^{+}  X_{2}(4013)$ and $pp \to \Lambda_{c}^{+}\Lambda_{c}^{+} X_{2}(4013)$ within an effective Lagrangian framework. The production processes are modeled via $t$-channel $D^*/\bar{D}^{*}$ meson exchanges, while initial-state interactions (ISI) mediated by Pomeron and Reggeon exchanges are also taken into account. Our calculations indicate that the total cross sections can reach the pb level, suggesting that $X_{2}(4013)$ may be accessible at current and future experiments such as AMBER@CERN and LHCb. Inclusion of ISI enhances the cross sections by nearly one order of magnitude. The differential distributions show distinct angular behaviors for the two reactions: the $K^{-}p$ reaction exhibits a forward-peaked distribution, whereas the $pp$ reaction shows a dip near central angles. This study provides a quantitative theoretical benchmark for future experimental searches of $X_2(4013)$ and highlights the importance of initial-state interactions (ISI) in high-energy particle investigations.
\end{abstract}

\maketitle
\section{INTRODUCTION}
In recent years, the discovery of numerous hadrons containing heavy quarks has dramatically expanded our understanding of strong interactions, driven by higher-energy experiments, improved statistics, and rapid advances in detection and analysis techniques~\cite{Brambilla:2019esw, Chen:2021ftn, Liu:2023hhl, Jia:2023upb, Brambilla:2010cs, Johnson:2024omq}. Many of these states lie beyond the scope of the conventional quark model, offering unique insights into the non-perturbative regime of quantum chromodynamics (QCD)~\cite{Liu:2024uxn}. Among the various theoretical frameworks proposed to interpret near-threshold exotic states, hadronic molecular states---bound or resonant systems of two or more conventional hadrons---have emerged as particularly compelling candidates~\cite{Liu:2024uxn}.

A prototypical molecular state candidate is the charmonium-like $X(3872)$, which was first observed by the Belle Collaboration in 2003 in the $J/\psi \pi^+\pi^-$ invariant mass spectrum from $B \to K J/\psi \pi^+\pi^-$ decays~\cite{Belle:2003nnu}, and subsequently confirmed by BaBar~\cite{BaBar:2004iez}, CDF~\cite{CDF:2003cab}, D0~\cite{D0:2004zmu}, CMS~\cite{CMS:2013fpt}, LHCb~\cite{LHCb:2011zzp}, and BESIII~\cite{BESIII:2013fnz}. Notably, later measurements observed $X(3872)$ not only in the $J/\psi \pi^+\pi^-$ channel but also in several other decay modes, including $\bar{D}^0 D^0 \pi^0$~\cite{Belle:2006olv}, $\bar{D}^{*0} D^0$~\cite{BESIII:2020nbj,BaBar:2007cmo,Belle:2008fma}, $J/\psi \gamma$~\cite{BaBar:2006fjg,BaBar:2008flx,Belle:2005lfc}, and $J/\psi \omega$~\cite{BESIII:2019qvy}. These observations can be naturally interpreted within a molecular framework. In particular, the $\bar{D}^0 D^0 \pi^0$~\cite{Belle:2006olv} and $\bar{D}^{*0} D^0$~\cite{BESIII:2020nbj,BaBar:2007cmo,Belle:2008fma} channels are dominated by tree-level decays of the $D\bar{D}^{*}$ molecular component, while pronounced isospin-violating effects further support interpreting $X(3872)$ as a $D\bar{D}^{*}$ molecular state with quantum numbers $J^{PC}=1^{++}$~\cite{Tornqvist:2004qy}. Its decays are primarily realized through loop processes involving the $D\bar{D}^{*}$ molecular component.

In addition to direct searches for the molecular nature of $X(3872)$, indirect approaches can also provide valuable insights. Interpreting $X(3872)$ as a $D\bar{D}^{*}$ molecular state, heavy quark symmetry~\cite{Neubert:1993mb} predicts the existence of its molecular partners, a $D^{*}\bar{D}^{*} $ state with spin-parity $J^{PC}=2^{++}$, denoted as $X_2(4013)$. Observing $X_2(4013)$ would offer a crucial test of whether $X(3872)$ is indeed a genuine $D\bar{D}^{*}$ molecular state. Beyond heavy quark symmetry, numerous theoretical studies also support the existence of the $D^{*}\bar{D}^{*}$ molecular state~\cite{Liu:2024uxn,Trunin:2016uks,LHCb:2016inz,Blumenhagen:2023abk,Nieves:2012tt,Guo:2013sya,Baru:2016iwj,Mehen:2011yh,Cheng:2018mkc,Alexandrou:2020mds,Molina:2009ct}. These studies cover several widely used theoretical frameworks---including contact-range effective field theory~\cite{Mehen:2011yh}, one-boson-exchange models~\cite{Cheng:2018mkc}, and lattice QCD~\cite{Alexandrou:2020mds}---and consistently predict its mass to be around 4013~MeV.

Despite strong theoretical support, experimental confirmation of $X_2(4013)$ remains lacking. In particular, direct analyses of the $D^{*}\bar{D}^{*}$ invariant mass spectrum have not revealed a clear resonance near 4013 MeV~\cite{DEramo:2021psx}. Although BESIII has reported potential signals in the processes $e^+ e^- \to \pi^+ \pi^- J/\psi$ and $e^+ e^- \to \pi^+ \pi^- h_c$~\cite{BESIII:2016bnd}, higher-statistics data are required for definitive confirmation. The proximity of $X_2(4013)$ to $Z_c(4020)$ suggests that complex coupled-channel dynamics may affect its observation~\cite{Liu:2020tqy}, highlighting the need for novel experimental strategies. Indeed, several alternative approaches have been explored, including kinematic reconstruction of $B^+ \to K^+ X_2(4013) \to K^+ \bar{D}^0 D^0 \pi^0$ to search for $X_2(4013)$~\cite{Wu:2023rrp}; however, isospin-breaking effects in this process may make it difficult to occur in practice. Radiative transitions such as $Y(4360) \to X_2(4013)\gamma$~\cite{Liu:2024ziu} have also been considered for direct detection of $X_2(4013)$. Direct $e^+ e^-$ collision searches can only set upper limits on the production cross section~\cite{BESIII:2019tdo}, indicating that more sensitive and diverse production mechanisms are required.

Kaon-induced reactions at facilities such as OKA@U-70~\cite{Obraztsov:2016}, SPS@CERN~\cite{Velghe:2016}, and the newly commissioned AMBER@CERN~\cite{Quintans:2022utc}, as well as proton-proton collisions at LHC@CERN~\cite{ATLAS:2012yve,CMS:2013btf}, ALICE@LHC~\cite{ALICE:2008ngc}, and RHIC@BNL~\cite{PHENIX:2001vgc}, offer promising experimental environments for the discovery of such exotic states. Future high-luminosity programs at colliders like the FCC~\cite{FCC:2018vvp} further enhance these prospects. Motivated by these opportunities, we study the production cross sections of $X_2(4013)$ in the reactions $K^- p \to  D_s^-\Lambda_c^+ X_2(4013)$ and $pp \to \Lambda_c^+ \Lambda_c^+ X_2(4013)$, incorporating initial-state interactions (ISI) of the $K^- p$ and $pp$ systems. This approach establishes theoretical benchmarks essential for guiding forthcoming experimental searches and maximizes the potential for observing $X_2(4013)$ across multiple channels.

 This paper is organized as follows: In Sec.~\ref{Sec: formulism}, we
present the theoretical formalism. In Sec.~\ref{sec:results}, the numerical
result and discussions are given, followed by conclusions in
the last section.

\section{THEORETICAL FORMALISM}\label{Sec: formulism}
\begin{figure}[htbp]
	\centering
	\subfigure[]
	{\begin{minipage}[t]{0.48\columnwidth}
			\centering
			\includegraphics[width=4.2cm, height=2cm]{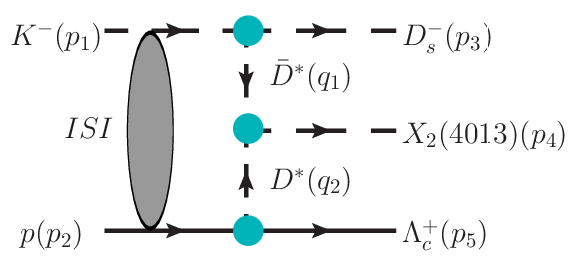}
			\label{1-1}
		\end{minipage}
	}
	\subfigure[]
	{\begin{minipage}[t]{0.48\columnwidth}
			\centering
			\includegraphics[width=4.2cm, height=2cm]{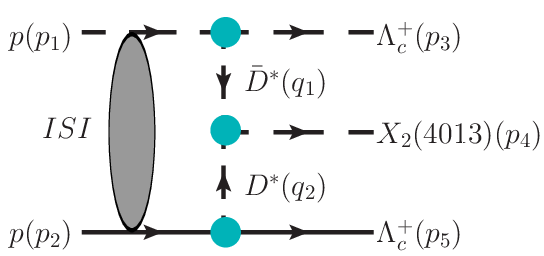}
			\label{1-2}
		\end{minipage}
	}
	\caption{ Tree-level Feynman diagrams for $X_2(4013)$ production in the $K^- p \to D_s^- \Lambda_c^+  X_2(4013)$ and $pp \to \Lambda_c^+ \Lambda_c^+ X_2(4013)$ reactions. These processes proceed via the exchange of $D^*$ and $\bar{D}^*$ mesons. The grey-shaded region indicates initial-state interactions (ISI). Particle momenta $(p_1, p_2, p_3, p_4, q_1, q_2)$ are indicated.} 
	\label{1}
\end{figure}

In this work, we focus on the production of $X_2(4013)$ in the reactions $K^- p \to D_s^-\Lambda_c^+  X_2(4013)$ and $pp \to \Lambda_c^+ \Lambda_c^+ X_2(4013)$. The corresponding tree-level Feynman diagrams are shown in Fig.~\ref{1}, which illustrate the central production mechanism. At high energies, the production is dominated by free $D^{*}$ and $\bar{D}^{*}$ mesons, whereas at lower energies, strong interactions between $D^{*}$ and $\bar{D}^{*}$ lead to the formation of the molecular state $X_2(4013)$.

To compute the amplitudes corresponding to Fig.~\ref{1}, we first define the effective Lagrangian densities for the relevant interaction vertices. The couplings for $X_2 \bar{D}^* D^*$, $\Lambda_c p D^*$, and $K \bar{D}^* D_s$ interactions are given by~\cite{Huang:2016ygf,Liu:2024ziu,Huang:2020ptc}:
\begin{align}
	\mathcal{L}_{X_2\bar{D}^*D^*} &= g_{X_2} X_2^{\mu\nu}(4013) D_\mu^* \bar{D}_\nu^*, \label{eq:L_X} \\
	\mathcal{L}_{\Lambda_c p D^*} &= g_{\Lambda_c p D^*} \bar{\Lambda}_c \gamma^\mu p D_\mu^{*0} + \text{H.c.}, \label{eq:L_Lambda} \\
	\mathcal{L}_{K\bar{D}^*D_s} &= i g_{K\bar{D}^*D_s} \bar{D}^{*\mu} \left[ \bar{D}_s \partial_\mu K - (\partial_\mu \bar{D}_s) K \right] + \text{H.c.}. \label{eq:L_K}
\end{align}
Here, the coupling $g_{\Lambda_c p D^*} = -5.20~\mathrm{GeV}$ is derived from SU(4) symmetry using benchmark values $g_{\pi NN} = 13.45~\mathrm{GeV}$ and $g_{\rho NN} = 6.0~\mathrm{GeV}$~\cite{Okubo:1975sc}. The couplings $g_{X_2} = 9.0~\mathrm{GeV}$ and $g_{K\bar{D}^*D_s} = 5.0~\mathrm{GeV}$ are adopted from Refs.~\cite{Liu:2024ziu,Huang:2020ptc}.

For the diagram in Fig.~\ref{1} (b), the $\Lambda_c p \bar{D}^*$ vertex is described by the Lagrangian~\cite{Huang:2016ygf}:
\begin{align}
	\mathcal{L}_{\Lambda_c p \bar{D}^*} = g_{\Lambda_c p \bar{D}^*} \bar{\Lambda}_c \gamma^\mu p \bar{D}_\mu^{*0} + \text{H.c.}, \label{eq:L_Lambdabar}
\end{align}
where $g_{\Lambda_c p \bar{D}^*} = -5.20~\mathrm{GeV}$ is fixed similarly through SU(4) symmetry.

Since hadrons are not pointlike, we include form factors for the exchanged $\bar{D}^*$ and $D^*$ mesons. We adopt a commonly used monopole form factor:
\begin{align}
	\mathcal{F}_i = \frac{\Lambda_i^2 - m_i^2}{\Lambda_i^2 - q_i^2}, \quad i = \bar{D}^*, D^*,
\end{align}
where $q_i$ and $m_i$ are the four-momentum and mass of the exchanged meson, respectively. The cutoff $\Lambda_i$ characterizes the hadronic size scale and is empirically chosen to exceed the meson mass by several hundred MeV, $\Lambda_i = m_i + \alpha \Lambda_{\mathrm{QCD}}$, with $\Lambda_{\mathrm{QCD}} = 220~\mathrm{MeV}$. The parameter $\alpha$, reflecting nonperturbative QCD effects at low energies, is treated as a free parameter to be discussed in Section~\ref{sec:results}.

The propagators for the $\bar{D}^*$ and $D^*$ mesons are given by
\begin{align}
	G_i^{\mu\nu}(q) = \frac{i \left(-g^{\mu\nu} + q_i^\mu q_i^\nu / m_i^2 \right)}{q_i^2 - m_i^2}, \quad i = \bar{D}^*, D^*,
\end{align}
where $\mu$ and $\nu$ denote the polarization indices of the corresponding vector mesons.

Using the effective Lagrangians and propagators introduced above, the invariant scattering amplitudes for the reactions shown in Fig.~\ref{1} can be written as
\begin{align}
	\mathcal{M}^{\text{Born}}_a &= i g_a \, \bar{u}(p_5) \gamma^\mu u(p_2) 
	\frac{-g^{\mu\nu} + q_2^\mu q_2^\nu / m_{D^*}^2}{q_2^2 - m_{D^*}^2} \, \epsilon^{\nu\sigma}(p_4) \nonumber \\
	&\quad \times \frac{-g^{\sigma\eta} + q_1^\sigma q_1^\eta / m_{\bar{D}^*}^2}{q_1^2 - m_{\bar{D}^*}^2} \, (p_3^\eta + p_1^\eta) \, \mathcal{F}, 
	\label{eq6} \\
	\mathcal{M}^{\text{Born}}_b &= - g_b \, \bar{u}(p_5) \gamma^\mu u(p_2) 
	\frac{-g^{\mu\nu} + q_2^\mu q_2^\nu / m_{D^*}^2}{q_2^2 - m_{D^*}^2} \, \gamma^\sigma \nonumber\\
	&\quad \times\frac{-g^{\sigma\eta} + q_1^\sigma q_1^\eta / m_{\bar{D}^*}^2}{q_1^2 - m_{\bar{D}^*}^2} \, \mathcal{F}, 
	\label{eq7}
\end{align}
where $g_a = g_{X_2} g_{\Lambda_c p D^*} g_{K \bar{D}^* D_s}$, $g_b = g_{X_2} g_{\Lambda_c p \bar{D}^*} g_{K \bar{D}^* D_s}$, and $\mathcal{F} = \mathcal{F}_{\bar{D}^*} \, \mathcal{F}_{D^*}$ denotes the product of the form factors for the exchanged mesons.

To enhance the predictive reliability, we incorporate initial-state interactions (ISI) in the $K^- p \to K^- p$ and $pp \to pp$ channels, which are represented by the shaded region in Fig.~\ref{1}. These ISI mechanisms are essential for a realistic description of the production dynamics. Our approach follows the simplified framework of Refs.~\cite{Lebiedowicz:2012, Lebiedowicz:2013vya}, where the elastic $\bar{K} N \to \bar{K} N$ and $NN \to NN$ amplitudes are modeled solely through Pomeron and Reggeon exchanges. Despite its simplicity, this model reproduces elastic scattering data with high accuracy~\cite{Lebiedowicz:2012, Lebiedowicz:2013vya}. By employing the optical theorem, the elastic amplitude can be consistently related to the total cross section, ensuring a coherent description of both elastic and total scattering.

\begin{figure}[htbp]
	\centering  
	\includegraphics[width=0.7\linewidth]{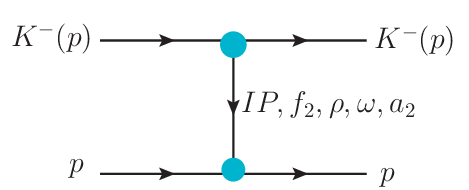}
	\caption{Initial-state interaction (ISI) mechanisms for the reactions $K^{-}p \to K^{-}p$ and $pp \to pp$.}
	\label{2-isi}
\end{figure}
The tree-level Feynman diagrams in Fig.~\ref{2-isi} provide a detailed illustration of these fundamental ISI exchange mechanisms, which include contributions from the Pomeron and the Reggeons \(f_2, a_2, \rho,\) and \(\omega\).
The full scattering amplitude for the $\bar{K}N \to \bar{K}N$ and $NN \to NN$ processes is obtained as a coherent sum of these exchanges~\cite{Lebiedowicz:2012,Lebiedowicz:2013vya}:
\begin{align}
	T(s,t) &= A_{\mathrm{IP}}(s,t) + A_{f_2}(s,t) \pm A_{a_2}(s,t) \nonumber\\
	&\quad + A_{\omega}(s,t) \pm A_{\rho}(s,t)\label{eq9},
\end{align}
where \(s\) and \(t\) are the squared center-of-mass energy and squared four-momentum transfer, respectively, and the energy scale is set as $s_0 = 1~\mathrm{GeV}^2$. The signs correspond to the specific charge channels: the upper signs apply to $K^-p$ or $pp$ scattering, while the lower signs apply to $K^-n$ scattering.

In the high-energy regime, each partial amplitude shown in Eq.~\ref{eq9} contributing to $\bar{K}N \to \bar{K}N$ or $NN \to NN$ scattering is parametrized using the standard Regge pole form~\cite{Lebiedowicz:2012,Lebiedowicz:2013vya}:
\begin{equation}
	A_i(s, t) = \eta_i \, s \, C^{\bar{K}N/NN}_i \left( \frac{s}{s_0} \right)^{\alpha_i(t)-1} \exp\left( \frac{B^{\bar{K}N/NN}_i}{2} t \right),
	\label{eq:Ai}
\end{equation}
where $\alpha_i(t) = \alpha_i(0) + \alpha_i' t$ is the linear Regge trajectory, $\eta_i$ is the signature factor, and $C^{\bar{K}N}_i$ ($C^{NN}_i$) is the residue coupling at the $\bar{K}N$ ($NN$) vertex. The slope parameters $B^{\bar{K}N/NN}_i$, which characterize the diffractive cone for each trajectory, are adopted from Refs.~\cite{Lebiedowicz:2012,Lebiedowicz:2013vya} as follows: 
$B^{\bar{K}N}_{\mathrm{IP}} = 5.5~\mathrm{GeV}^{-2}$, $B^{\bar{K}N}_{\mathrm{IR}} = 4.0~\mathrm{GeV}^{-2}$, $B^{NN}_{\mathrm{IP}} = 9.0~\mathrm{GeV}^{-2}$, and $B^{NN}_{\mathrm{IR}} = 6.0~\mathrm{GeV}^{-2}$. 
Here, IR collectively denotes the Regge intercepts $f_2, a_2, \rho, \omega$. The remaining parameters--- $C^{\bar{K}N}_i, C^{NN}_i, \alpha_i(0), \alpha_i'$ and $\eta_i$---are taken from Ref.~\cite{Lebiedowicz:2012} and listed in Table~\ref{tab2}. This complete parameter set is constrained by fits to experimental data for both elastic differential and total cross sections. 
\begin{table}[htbp]
	\centering
	\caption{The parameters of the Pomeron and Reggeon exchanges were determined based on elastic and total cross section data in Refs.~\cite{Lebiedowicz:2012,Lebiedowicz:2013vya}.}
	\begin{tabular}{lcc r@{.}l r@{.}l}
		\hline\hline    $i$&~~~$\eta_i$&~~~$\alpha_i(t)$&\multicolumn{2}{l}{$C_i^{\bar{K}N}$(mb)}&		\multicolumn{2}{l}{$C_i^{NN}$(mb)} \\  
		\hline
		$IP$&~~~$i$&~~~$1.081+(0.25 \,\text{GeV}^{-2})$t&~~~11&82&21&70 \\
		$f_2$&~~~$-0.861+i$&~~~$0.548+(0.93\,\text{GeV}^{-2})$t&~~~15&67&75&4875\\
		$\rho$&~~~$-1.162-i$&~~~$0.548+(0.93\,\text{GeV}^{-2})$t&~~~2&05&1&0925\\
		$\omega$&~~~$-1.162-i$&~~~$0.548+(0.93\,\text{GeV}^{-2})$t&~~~7&055&20&0625\\
		$a_2$&~~~$-0.861+i$&~~~$0.548+(0.93\,\text{GeV}^{-2})$t&~~~1&585&1&7475\\
		\hline\hline
		
	\end{tabular}
	\label{tab2}
\end{table}

After incorporating the initial-state interactions, the full amplitude can be written as~\cite{Lebiedowicz:2012}:
\begin{align}
	{\cal M}^{\rm full} &= {\cal M}^{\rm Born} + \frac{i}{16\pi^2 s} \int d^2 \vec{k}_t\, \mathcal{T}(s, k_t^2)\, {\cal M}^{\rm Born}(s, k_t^2),
\end{align}
where \(\vec{k}_t\) denotes the momentum transfer in the \(K^- p \to K^- p\) or \(pp \to pp\) reaction.  Using this amplitude, we calculate the differential cross sections for the processes \(K^- p \to  D_s^-\Lambda_c^+ X_2(4013)\) and \(pp \to \Lambda_c^+ \Lambda_c^+ X_2(4013)\) in the center-of-mass frame. Since these are \(2\to3\) reactions, the cross sections follow from the total amplitude according to
\begin{align}
	d\sigma &= \frac{m_p}{2\sqrt{(p_1 \cdot p_2)^2 - m_1^2 m_2^2}} \sum_{s_i, s_f} |{\cal M}^{\rm full}|^2 \nonumber\\
	&\quad \times \frac{d^3 \vec{p}_3}{2 E_3} \frac{d^3 \vec{p}_4}{2 E_4} \frac{m_{\Lambda_c} d^3 \vec{p}_5}{E_5} \delta^4(p_1 + p_2 - p_3 - p_4 - p_5),
\end{align}
Here, \(E_3, E_4\), and \(E_5\) denote the energies of \(D_s^-,\, X_2\), \(\Lambda_c^+\) in the \(K^- p\) reaction, and of \(\Lambda_c^+,\, X_2\), \(\Lambda_c^+\) in the \(pp\) reaction, respectively. The employed particle masses are  \(m_{K^-} = 493.68~\text{MeV}\), \(m_{\Lambda_c}= 2286.46~\text{MeV}\), and \(m_p = 938.27~\text{MeV}\).

\section{Results and Discussions}\label{sec:results}
In this work, our main objective is to search for the theoretically predicted \(J^{PC}=2^{++}\) \(\bar{D}^{*}D^{*}\) molecular state, for which only tentative experimental indications exist but no firm confirmation has yet been achieved~\cite{BESIII:2016bnd}.  We propose to study its production through the reactions 
\(K^-p \rightarrow D_s^- \Lambda_c^+  X_2(4013)\) and 
\(pp \rightarrow \Lambda_c^+ \Lambda_c^+ X_2(4013)\). 
The calculations are performed within the effective Lagrangian framework, considering the central production mechanism~\cite{Lebiedowicz:2012, Lebiedowicz:2013vya} with \(t\)-channel exchange of \(D^{*}\) and \(\bar{D}^{*}\) mesons.  The $s$- and $u$-channel contributions are not considered because they would necessitate the production of baryons containing two $c\bar{c}$ pairs. The corresponding thresholds are quite high, with the lowest possible masses being $m_{D_s^-}+m_{X_2(4013)}+m_{\Lambda_c^+}=8267.81~\text{MeV}$ for the reaction $K^{-}p\to D_s^{-}\Lambda_c^{+}X_2(4013)$, and $m_{\Lambda_c^+}+m_{X_2(4013)}+m_{\Lambda_c^+}=8585.92~\text{MeV}$ for $pp\to \Lambda_c^{+}\Lambda_c^{+}X_2(4013)$. 


The dominant source of theoretical uncertainty arises from the parameter $\alpha$, which characterizes the form factor associated with $D^{*}$ and $\bar{D}^{*}$ mesons exchange. Since $\alpha$ cannot be determined from first principles, it is constrained phenomenologically through fits to experimental data. Fortunately, values of $\alpha=1.5$ and 1.7 have been obtained by fitting the processes $e^+e^- \rightarrow D\bar{D}$\cite{Belle:2007qxm} and $e^+e^- \rightarrow \gamma_{\rm ISR} D\bar{D}$\cite{BaBar:2006qlj}, as detailed in Ref.~\cite{Guo:2016iej}. In the present analysis, we adopt $\alpha=1.5$ and 1.7 for our calculations.

\begin{figure}[htbp]
	\begin{minipage}{\columnwidth}
		\centering  
		\includegraphics[width=\linewidth]{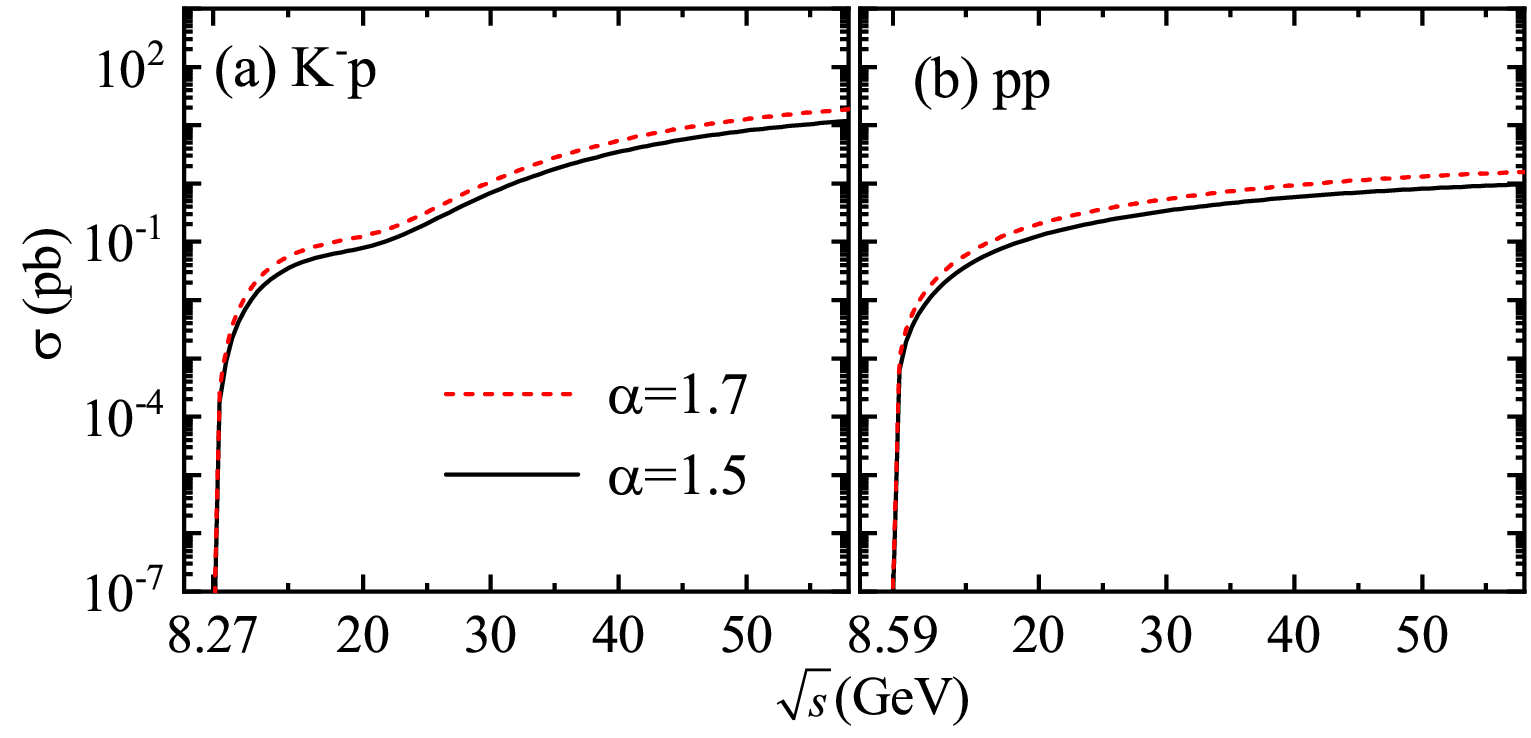}
		\caption{The total cross sections for the \(K^-p\rightarrow D_s^-\Lambda_c^+X_2(4013)\) and \(pp\rightarrow \Lambda_c^+\Lambda_c^+X_2(4013)\) reactions are evaluated for different $\alpha$ values. }\label{2}
	\end{minipage}
\end{figure}
Using the determined $\alpha$ values, the cross sections for the reactions $K^{-}p \to D_{s}^{-} \Lambda_{c}^{+}X_{2}(4013)$ and $pp \rightarrow \Lambda_{c}^{+} \Lambda_{c}^{+}X_{2}(4013)$ are evaluated. The theoretical results, obtained with a cutoff $\alpha = 1.5$ or $1.7$, are shown in Fig.~\ref{2} for energies from near threshold up to 58~GeV. The cross sections exhibit a sharp increase near the thresholds, occurring at 8267.81~MeV and 8585.92~MeV for the two reactions, respectively, due to the opening of phase space. Beyond the thresholds, the cross sections continue to rise more gradually and eventually stabilize at higher energies.  Moreover, the production cross section of $X_2(4013)$ in the 
$K^- p$ reaction is larger than that in the $pp$ reaction. 
For example, at 50.27 GeV, the \(K^-p\) collision cross section stabilizes at 8.161 pb (\(\alpha= 1.5\)) and 12.881 pb (\(\alpha= 1.7\)), while at 50.09 GeV, the \(pp\) collision cross section stabilizes at 0.813 pb (\(\alpha\) = 1.5) and 1.315 pb (\(\alpha= 1.7\)).

The results also show that the line shapes of the cross sections for the 
$K^- p \to  D_s^-\Lambda_c^+ X_2(4013)$ and 
$pp \to \Lambda_c^+ \Lambda_c^+ X_2(4013)$ reactions 
differ significantly in the energy region 
$\sqrt{s}=20.77$--$36.77$~GeV. 
That is, in this region, the cross section for the 
$K^- p \to D_s^- \Lambda_c^+X_2(4013)$ reaction increases 
much more rapidly than that for the 
$pp \to \Lambda_c^+ \Lambda_c^+ X_2(4013)$ reaction.
This difference may arise from the significant role of the Pauli exclusion principle in the $pp$ initial state.
 In the $pp$ reaction, the cross section rises smoothly above threshold due to the competition between short-range attraction and Pauli repulsion. At sufficiently high energies, the dynamics of internal quarks and gluons become dominant, leading to a saturation of the cross section. In contrast, the $K^- p$ reaction is free from Pauli constraints, allowing wavefunction overlap under interaction, which enhances the production cross section (corresponding to \(\sqrt{s}=20.77-36.77~\mathrm{GeV}\)). At even higher energies, the reaction gradually becomes dominated by quark and gluon dynamics, and the cross section saturates in this region as well.  Moreover, the production of the $\Lambda_c^+ \Lambda_c^+$ pair in the $pp$ reaction is also subject to Pauli exclusion effects, which may partially explain why the $X_2(4013)$ production cross section in $pp$ collisions is smaller than that in $K^- p$ reactions.

As shown in Fig.~\ref{2}, the cross sections exhibit a strong sensitivity to the form factor parameter \(\alpha\).  For the reaction \(K^- p \to  D_s^-\Lambda_c^+ X_2(4013)\) at \(\sqrt{s}=30.27\ \mathrm{GeV}\), we obtain
\[
\sigma(\alpha=1.5) \approx 0.728~\mathrm{pb}, \qquad 
\sigma(\alpha=1.7) \approx 1.139~\mathrm{pb}.
\]
Hence the enhancement factor is
\[
R \equiv \frac{\sigma(\alpha=1.7)}{\sigma(\alpha=1.5)} \approx \frac{1.139}{0.728} \approx 1.564,
\]
which indicates that the cross section for \(\alpha=1.7\) is roughly \(56.4\%\) larger than that for \(\alpha=1.5\). 
For the $pp \to \Lambda_c^+ \Lambda_c^+ X_2(4013)$ reaction at \(\sqrt{s}=30.09\,\text{GeV}\), the $\alpha$-dependent factor $R$ is 1.598. 
However, as the energy increases, this factor becomes larger. For example, we obtain $R=1.578$ for the $K^- p \to D_s^-\Lambda_c^+ X_2(4013)$ reaction at $\sqrt{s}=50.27~\text{GeV}$ and $R=1.617$ for the $pp \to \Lambda_c^+ \Lambda_c^+ X_2(4013)$ reaction at $\sqrt{s}=50.09~\text{GeV}$. This result further indicates that particular caution must be exercised when treating the form factor of the \(D^*\) and \(\bar{D}^*\) mesons, and highlights the importance of precise experimental determination of this parameter.

\begin{figure}[http]
	\begin{minipage}{\columnwidth}
		\centering  
		\includegraphics[width=\linewidth]{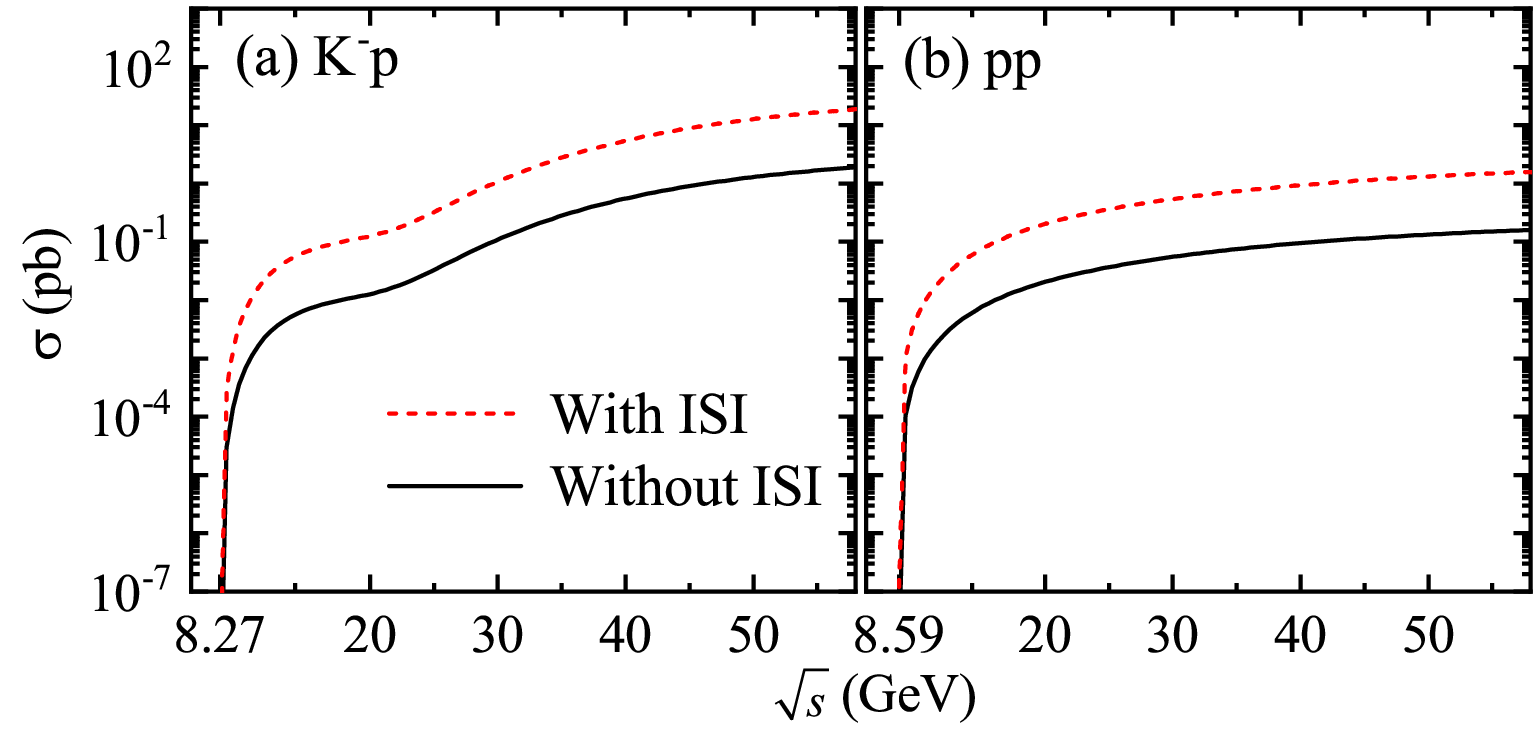}
		\caption{The cross section with and without ISI for the \(K^-p\rightarrow D_s^-\Lambda_c^+X_2(4013)\) and \(pp\rightarrow \Lambda_c^+\Lambda_c^+X_2(4013)\) reactions with $\alpha=1.7$.  }\label{3}
	\end{minipage}
\end{figure}

To illustrate the role of initial-state interactions (ISI) in the production of \(X_2(4013)\), we compare the cross sections for \(K^-p\) and \(pp\) collisions with the cutoff parameter fixed at \(\alpha = 1.7\). Figure~\ref{3}(a) shows the result including \(K^-p\) ISI, while Fig.~\ref{3}(b) presents the corresponding case for \(pp\) ISI.  In both reactions, ISI leads to a pronounced enhancement of the cross section. For instance, at \(\sqrt{s}=50.27~\text{GeV}\), the total cross section for \(K^-p\) scattering is \(12.881~\text{pb}\), compared with the Born-level value of \(1.304~\text{pb}\), yielding an enhancement factor of about 9.878. 
Likewise, at \(\sqrt{s}=50.09~\text{GeV}\), the \(pp\) cross section increases from \(0.133~\text{pb}\) to \(1.315~\text{pb}\), corresponding to a factor of roughly 9.887.  Although the enhancement factors are nearly identical, the effect in \(pp\) collisions is marginally larger. This subtle difference may stem from the intrinsic distinctions in ISI dynamics between meson–baryon and baryon–baryon systems~\cite{Garcilazo:2022edi}.  Such modifications, arising from the $K^{-}p$ and $pp$ interactions, indicate that these effects are essential and cannot be neglected in studies of heavy-quark particle production.

\begin{figure}[http]
 	\begin{minipage}{\columnwidth}
 		\centering  
 		\includegraphics[width=1.0\linewidth]{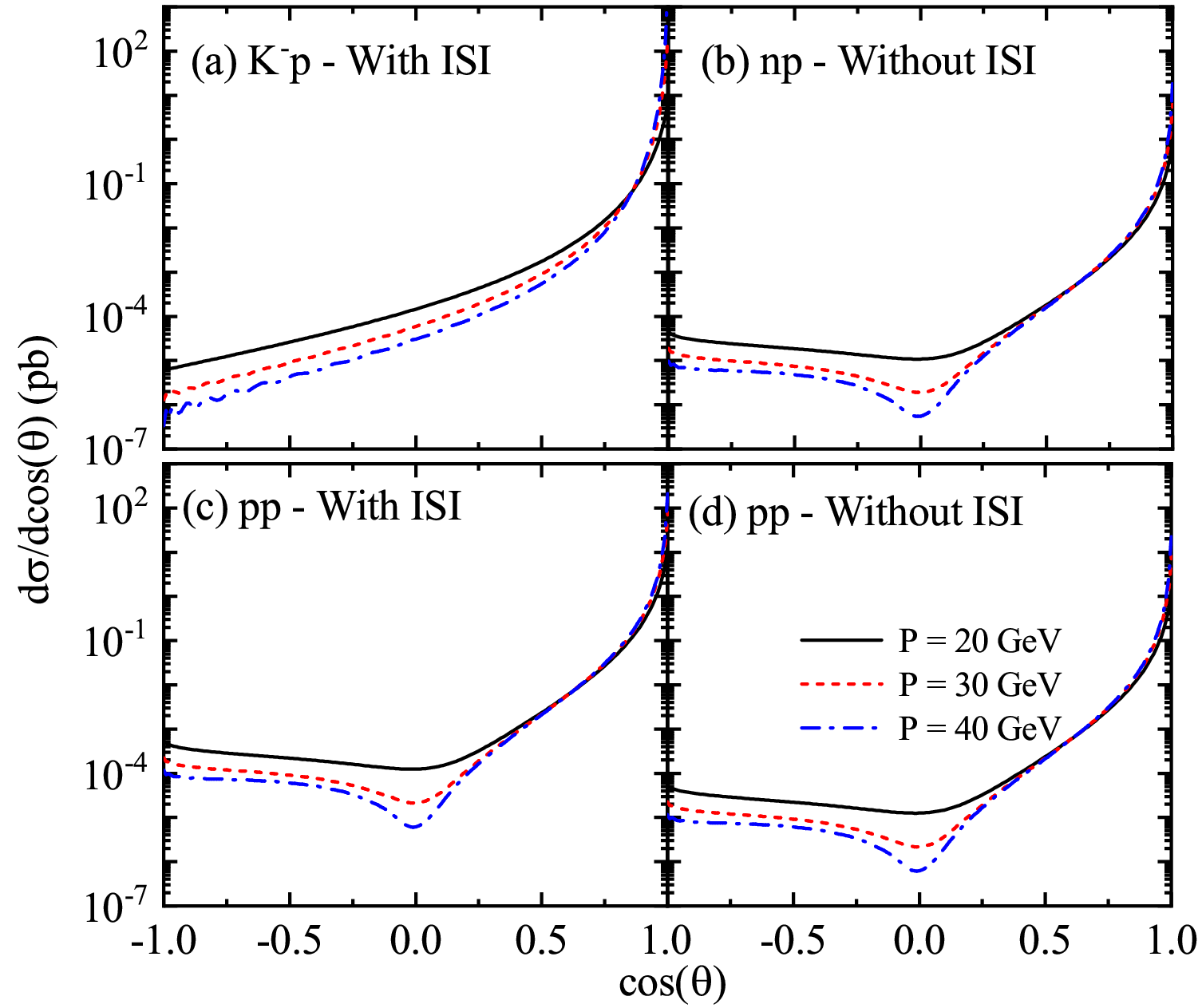}
 		\caption{Differential cross sections for (a) \(K^-p\rightarrow D_s^-\Lambda_c^+X_2(4013)\) (with ISI), (b) \(np\rightarrow \Lambda_c^+ \Sigma_c(2445) X_2(4013)\), and (c)(d) \(pp\rightarrow \Lambda_c^+\Lambda_c^+X_2(4013)\) (with and without ISI, respectively), calculated with $\alpha=1.7$. 
 		The black solid lines, red dashed lines, and blue dashed-dotted lines are obtained at beam energies \(P=20\), 30, and 40 GeV, respectively. }\label{4}
 	\end{minipage}
 \end{figure}

In addition to the total cross sections, we also calculated the differential cross sections for the reactions 
\(K^-p\rightarrow D_s^- \Lambda_c^+  X_2(4013)\) and 
\(pp\rightarrow \Lambda_c^+ \Lambda_c^+ X_2(4013)\) as functions of the scattering angle of the outgoing 
$\Lambda_c$ relative to the beam direction at different kaon and proton beam energies, 
i.e., \(P_{K^- / p} = 20.0\), 30.0, and 40.0~GeV. The theoretical results are shown in Fig.~\ref{4}.

As shown in Fig.~\ref{4}, for the reaction 
\(K^-p \rightarrow D_s^-\Lambda_c^+  X_2(4013)\), 
the differential cross section exhibits a simple, monotonic behavior: it is strongly enhanced 
in the extreme forward direction (\(\cos\theta \to 1\)) and decreases smoothly toward the 
backward direction (\(\cos\theta \to -1\)), which is a typical signature of a $t$-channel 
contribution. This behavior is consistent with the physical processes considered in our study 
(see Fig.~\ref{1}(a)). Interestingly, although the reaction mechanism for 
\(pp \rightarrow \Lambda_c^+ \Lambda_c^+ X_2(4013)\) is the same as that for 
\(K^-p \rightarrow D_s^- \Lambda_c^+ X_2(4013)\)---both being central production processes 
mediated by $t$-channel $D^*$ and $\bar{D}^{*}$ exchanges---the differential cross section for the 
\(pp\) reaction behaves quite differently. It does not display the characteristic features of 
a $t$-channel contribution; instead, it exhibits a non-monotonic shape, with a pronounced dip 
near central angles (\(\cos\theta \approx 0\)) and a significant increase toward both the 
forward and backward directions. This trend becomes more pronounced as the incident proton energy increases, as shown in Fig.~\ref{4}(c). Specifically, as the incident proton energy varies from 20 GeV to 40 GeV, the dip structure becomes increasingly prominent.

It should be noted that the dip structure observed in the differential cross section of the 
\(pp \rightarrow \Lambda_c^+ \Lambda_c^+ X_2(4013)\) reaction does not originate from the 
initial-state interaction (ISI) of the \(pp \rightarrow pp\) system. As shown in Fig.~\ref{4}(d), 
even when the ISI contribution is switched off, a similar dip still appears around 
\(\cos\theta = 0\).  At the same time, we have also ruled out the possibility that the dip is caused by the presence of identical $p$ and $\Lambda_c^+$ in the initial and final states, which would invoke the Pauli exclusion principle. We have further calculated the reaction $np \rightarrow \Lambda_c^+ \Sigma_c(2445) X_2(4013)$. The results show that a similar dip structure also appears in the differential cross section of $np \rightarrow \Lambda_c^+ \Sigma_c(2445) X_2(4013)$ (see Fig.~\ref{4}(b)).  One possible reason is that this dip arises from a "dynamical zero" caused by helicity conservation. The detailed differences can be seen in their respective amplitudes, given by Eq.~\ref{eq6} and Eq.~\ref{eq7}.  From this, we find that the $K^{-}p$ system involves a relatively simple meson derivative coupling, whereas the $pp$ system features a more complex baryon current coupling with nontrivial helicity structure. The latter is subject to stronger constraints from helicity conservation when coupling to high-spin final-state particles, which gives rise to the distinctive central dip observed in the differential cross section, and this effect is dependent on the incident energy.

\section{Summary}
Thanks to significant experimental progress, it is now possible to search for the $J^P=2^{++}$ $D^{*}\bar{D}^{*}$ molecular state $X_2(4013)$, which is predicted by heavy-quark symmetry as the spin-2 partner of the $X(3872)$, commonly interpreted as a $D\bar{D}^{*}$ molecule. Despite previous efforts, we propose to investigate it via the reactions $K^-p \rightarrow D_s^- \Lambda_c^+  X_2(4013)$ and $pp \rightarrow \Lambda_c^+ \Lambda_c^+ X_2(4013)$. The reaction mechanisms mainly involve central production through $t$-channel $D^{*}$ and $\bar{D}^{*}$ exchange, while also accounting for initial-state interactions (ISI) dominated by Pomeron and Reggeon exchanges. The only parameter in the model is the form factor $\alpha$, for which we adopt the values $\alpha = 1.5$ and $1.7$, determined from fits to experimental data.

The calculated total cross sections for both reaction channels can reach the pb level, suggesting that experimental observation is feasible at facilities such as AMBER@CERN and LHCb. Specifically, at a center-of-mass energy of 50.27~GeV, the \(K^-p\) collision cross section stabilizes at 8.161~pb for \(\alpha = 1.5\) and 12.881~pb for \(\alpha = 1.7\). Meanwhile, at 50.09~GeV, the \(pp\) collision cross section stabilizes at 0.813~pb (\(\alpha = 1.5\)) and 1.315~pb (\(\alpha = 1.7\)).  The \(K^{-}p\) and \(pp\) initial-state interactions (ISI) mediated by Pomeron and Reggeon exchanges play a crucial role in the search for the \(X_{2}(4013)\), as they enhance the total cross sections by nearly an order of magnitude compared with the results without ISI.  The differential cross sections exhibit distinctly different angular behaviors: the \(K^-p\) reaction shows a monotonic forward-peaking distribution, typical of \(t\)-channel–dominated processes, whereas the \(pp\) reaction displays a non-monotonic dip near \(\cos\theta \approx 0\).

\section*{Acknowledgments}
This work was not supported by any funding agency.


\begin{thebibliography}{99}
	
	
	\bibitem{Brambilla:2019esw}
	N.~Brambilla, S.~Eidelman, C.~Hanhart, A.~Nefediev, C.~P.~Shen, C.~E.~Thomas, A.~Vairo and C.~Z.~Yuan,
	The $XYZ$ states: experimental and theoretical status and perspectives,
	\href{Phys. Rept. 873, 1-154 (2020)}
	{Phys. Rept. \textbf{873}, 1-154 (2020)}.
	
	
	\bibitem{Chen:2021ftn}
	S.~Chen, Y.~Li, W.~Qian, Z.~Shen, Y.~Xie, Z.~Yang, L.~Zhang and Y.~Zhang,
	Heavy Flavour Physics and CP Violation at LHCb: a Ten-Year Review,
	\href{Front. Phys. 18, 44601 (2023)}
	{Front. Phys. \textbf{18}, 44601 (2023)}.
	
	
	\bibitem{Liu:2023hhl}
	Z.~Liu and R.~E.~Mitchell,
	New hadrons discovered at BESIII,
	\href{Sci. Bull. 68, 2148-2150 (2023)}{
	Sci. Bull. \textbf{68}, 2148-2150 (2023)}.
	
	
	\bibitem{Jia:2023upb}
	S.~Jia, W.~Xiong and C.~Shen,
	Status and Prospects of Exotic Hadrons at Belle II,
	\href{Chin. Phys. Lett. 40, no.12, 121301 (2023)}{
	Chin. Phys. Lett. \textbf{40}, no.12, 121301 (2023)}.
	
	
	\bibitem{Brambilla:2010cs}
	N.~Brambilla, S.~Eidelman, B.~K.~Heltsley, R.~Vogt, G.~T.~Bodwin, E.~Eichten, A.~D.~Frawley, A.~B.~Meyer, R.~E.~Mitchell and V.~Papadimitriou, \textit{et al.}
	Heavy Quarkonium: Progress, Puzzles, and Opportunities,
	\href{Eur. Phys. J. C 71, 1534 (2011)}{
	Eur. Phys. J. C \textbf{71}, 1534 (2011)}.
	
	
	\bibitem{Johnson:2024omq}
	D.~Johnson, I.~Polyakov, T.~Skwarnicki and M.~Wang,
	Exotic Hadrons at LHCb,
	\href{Ann. Rev. Nucl. Part. Sci. 74, 583-612 (2024) }{
	Ann. Rev. Nucl. Part. Sci. \textbf{74}, 583-612 (2024)}.
	
	\bibitem{Liu:2024uxn}
	M.~Z.~Liu, Y.~W.~Pan, Z.~W.~Liu, T.~W.~Wu, J.~X.~Lu and L.~S.~Geng,
	Three ways to decipher the nature of exotic hadrons: Multiplets, three-body hadronic molecules, and correlation functions,
	\href{Phys. Rept. 1108, 1-108 (2025)}{
	Phys. Rept. \textbf{1108}, 1-108 (2025)}.
	
	
	\bibitem{Belle:2003nnu}
	S.~K.~Choi \textit{et al.} [Belle],
	Observation of a narrow charmonium-like state in exclusive $B^\pm \to K^\pm \pi^+ \pi^- J/\psi$ decays,
	\href{ Phys. Rev. Lett. 91, 262001 (2003)}{
	Phys. Rev. Lett. \textbf{91}, 262001 (2003)}.
	
	
	\bibitem{BaBar:2004iez}
	B.~Aubert \textit{et al.} [BaBar],
	Observation of the decay $B \to J/\psi \eta K$ and search for $X(3872) \to J/\psi \eta$,
	\href{Phys. Rev. Lett. 93, 041801 (2004)}{
	Phys. Rev. Lett. \textbf{93}, 041801 (2004)}.
	
	
	\bibitem{CDF:2003cab}
	D.~Acosta \textit{et al.} [CDF],
	Observation of the narrow state $X(3872) \to J/\psi \pi^+ \pi^-$ in $\bar{p}p$ collisions at $\sqrt{s} = 1.96$ TeV,
	\href{Phys. Rev. Lett. 93, 072001 (2004)}{
	Phys. Rev. Lett. \textbf{93}, 072001 (2004)}.
	
	\bibitem{D0:2004zmu}
	V.~M.~Abazov \textit{et al.} [D0],
	Observation and properties of the $X(3872)$ decaying to $J/\psi \pi^+ \pi^-$ in $p\bar{p}$ collisions at $\sqrt{s} = 1.96$ TeV,
	\href{	Phys. Rev. Lett. 93, 162002 (2004)}{
	Phys. Rev. Lett. \textbf{93}, 162002 (2004)}.
	
	\bibitem{CMS:2013fpt}
	S.~Chatrchyan \textit{et al.} [CMS],
	Measurement of the $X$(3872) Production Cross Section Via Decays to $J/\psi \pi^+ \pi^-$ in $pp$ collisions at $\sqrt{s}$ = 7 TeV,
	\href{JHEP 04, 154 (2013)}{
	JHEP \textbf{04}, 154 (2013)}.
	
	\bibitem{LHCb:2011zzp}
	R.~Aaij \textit{et al.} [LHCb],
	Observation of $X(3872) $ production in $pp$ collisions at $\sqrt{s}=7$ TeV,
	\href{Eur. Phys. J. C 72, 1972 (2012)}{
	Eur. Phys. J. C \textbf{72}, 1972 (2012)}.
	
	\bibitem{BESIII:2013fnz}
	M.~Ablikim \textit{et al.} [BESIII],
	Observation of $e^+e^- \to \gamma X$(3872) at BESIII,
	\href{Phys. Rev. Lett. 112, no.9, 092001 (2014)}{
	Phys. Rev. Lett. \textbf{112}, no.9, 092001 (2014)}.
	
	\bibitem{Belle:2006olv}
	G.~Gokhroo \textit{et al.} [Belle],
	Observation of a Near-threshold 
	\(D^0\bar{D}^0\pi^0\) Enhancement in \(B\to D^0\bar{D}^0\pi^0K\)  Decay,
	\href{Phys. Rev. Lett. 97, 162002 (2006) }{
	Phys. Rev. Lett. \textbf{97}, 162002 (2006)}.
	
	\bibitem{BESIII:2020nbj}
	M.~Ablikim \textit{et al.} [BESIII],
	Study of Open-Charm Decays and Radiative Transitions of the $X(3872)$,
	\href{Phys. Rev. Lett. 124, no.24, 242001 (2020)}{
	Phys. Rev. Lett. \textbf{124}, no.24, 242001 (2020)}.
	
	\bibitem{BaBar:2007cmo}
	B.~Aubert \textit{et al.} [BaBar],
	Study of Resonances in Exclusive \(B\) Decays to \(\bar{D}^{(*)}D^{(*)}K\), 
	\href{Phys. Rev. D 77, 011102 (2008)}{
	Phys. Rev. D \textbf{77}, 011102 (2008)}.
	
	\bibitem{Belle:2008fma}
	T.~Aushev \textit{et al.} [Belle],
	Study of the \(B\to X(3872)(\to D^{*0}\bar{D}^0)K\) decay,
	\href{Phys. Rev. D 81, 031103 (2010)}{
	Phys. Rev. D \textbf{81}, 031103 (2010)}.
	
	\bibitem{BaBar:2006fjg}
	B.~Aubert \textit{et al.} [BaBar],
	Search for $B^{+} \to X(3872) K^{+}$, $X_{3872} \to J/\psi \gamma$,
	\href{Phys. Rev. D 74, 071101 (2006)}{
	Phys. Rev. D \textbf{74}, 071101 (2006)}.
	
	\bibitem{BaBar:2008flx}
	B.~Aubert \textit{et al.} [BaBar],
	Evidence for $X(3872) \to \psi_{2S} \gamma$ in $B^\pm \to X_{3872} K^\pm$ decays, and a study of $B \to c \bar{c} \gamma K$,
	\href{Phys. Rev. Lett. 102, 132001 (2009)}{
	Phys. Rev. Lett. \textbf{102}, 132001 (2009)}.
	
	\bibitem{Belle:2005lfc}
	K.~Abe \textit{et al.} [Belle],
	Evidence for \(X(3872)\to \gamma J/\psi\) and the sub-threshold decay \(X(3872)\to \omega J/\psi\),
	[arXiv:hep-ex/0505037 [hep-ex]].
	
	
	\bibitem{BESIII:2019qvy}
	M.~Ablikim \textit{et al.} [BESIII],
	Study of $e^+e^- \to \gamma \omega J/\psi$ and Observation of $X(3872) \to \omega J/\psi$,
	\href{Phys. Rev. Lett. 122, no.23, 232002 (2019) }{
	Phys. Rev. Lett. \textbf{122}, no.23, 232002 (2019)}.
	
	
	
	
	
	\bibitem{Tornqvist:2004qy}
	N.~A.~Tornqvist,
	Isospin breaking of the narrow charmonium state of Belle at 3872-MeV as a deuson,
	\href{Phys. Lett. B 590, 209-215 (2004)}{
	Phys. Lett. B \textbf{590}, 209-215 (2004)}.
	
	\bibitem{Neubert:1993mb}
	M.~Neubert,
	Heavy quark symmetry,
	\href{Phys. Rept. 245, 259-396 (1994)}{
	Phys. Rept. \textbf{245}, 259-396 (1994)}.
	
	\bibitem{Trunin:2016uks}
	A.~Trunin,
	\(bc\) diquark pair production in high energy proton-proton collisions,
	\href{Phys. Rev. D 93, no.11, 114029 (2016)}{
	Phys. Rev. D \textbf{93}, no.11, 114029 (2016)}.
	
	\bibitem{LHCb:2016inz}
	R.~Aaij \textit{et al.} [LHCb],
	Search for massive long-lived particles decaying semileptonically in the LHCb detector,
	\href{Eur. Phys. J. C 77, no.4, 224 (2017)}{
	Eur. Phys. J. C \textbf{77}, no.4, 224 (2017)}.
	
	\bibitem{Blumenhagen:2023abk}
	R.~Blumenhagen, C.~Kneissl and C.~Wang,
	Dynamical Cobordism Conjecture: solutions for end-of-the-world branes,
	\href{JHEP 05, 123 (2023)}{JHEP \textbf{05}, 123 (2023)}.
	
	
	
	
	
	\bibitem{Mehen:2011yh}
	T.~Mehen and J.~W.~Powell,
	Heavy Quark Symmetry Predictions for Weakly Bound B-Meson Molecules,
	\href{	Phys. Rev. D 84, 114013 (2011)}{
	Phys. Rev. D \textbf{84}, 114013 (2011)}.
	
	\bibitem{Cheng:2018mkc}
	L.~Cheng, O.~Eberhardt and C.~W.~Murphy,
	Novel theoretical constraints for color-octet scalar models,
	\href{Chin. Phys. C 43, no.9, 093101 (2019)}{
	Chin. Phys. C \textbf{43}, no.9, 093101 (2019)}.
	
	\bibitem{Alexandrou:2020mds}
	C.~Alexandrou, A.~Athenodorou, K.~Hadjiyiannakou and A.~Todaro,
	Neutron electric dipole moment using lattice QCD simulations at the physical point,
	\href{	Phys. Rev. D 103, no.5, 054501 (2021)}{
	Phys. Rev. D \textbf{103}, no.5, 054501 (2021)}.
	
	
	
	\bibitem{Nieves:2012tt}
	J.~Nieves and M.~P.~Valderrama,
	The Heavy Quark Spin Symmetry Partners of the \(X(3872)\),
	\href{Phys. Rev. D 86, 056004 (2012)}{
	Phys. Rev. D \textbf{86}, 056004 (2012)}.
	
	\bibitem{Guo:2013sya}
	F.~K.~Guo, C.~Hidalgo-Duque, J.~Nieves and M.~P.~Valderrama,
	Consequences of Heavy Quark Symmetries for Hadronic Molecules,
	\href{Phys. Rev. D 88, 054007 (2013)}{
	Phys. Rev. D \textbf{88}, 054007 (2013)}.
	
	\bibitem{Baru:2016iwj}
	V.~Baru, E.~Epelbaum, A.~A.~Filin, C.~Hanhart, U.~G.~Mei\ss{}ner and A.~V.~Nefediev,
	Heavy-quark spin symmetry partners of the \(X (3872)\) revisited,
	\href{Phys. Lett. B 763, 20-28 (2016)}{
	Phys. Lett. B \textbf{763}, 20-28 (2016)}.


 \bibitem{Molina:2009ct}
 R.~Molina and E.~Oset,
 The $Y(3940)$, $Z(3930)$ and the $X(4160)$ as dynamically generated resonances from the vector-vector interaction,
 \href{Phys. Rev. D 80, 114013 (2009)}{
 Phys. Rev. D \textbf{80}, 114013 (2009)}.

   
	

	
	\bibitem{DEramo:2021psx}
	F.~D'Eramo, F.~Hajkarim and S.~Yun,
	Thermal Axion Production at Low Temperatures: A Smooth Treatment of the QCD Phase Transition,
	\href{Phys. Rev. Lett. 128, no.15, 152001 (2022)}{
	Phys. Rev. Lett. \textbf{128}, no.15, 152001 (2022)}.
	
	
	
	\bibitem{BESIII:2016bnd}
	M.~Ablikim \textit{et al.} [BESIII],
	Precise measurement of the $e^+e^-\to \pi^+\pi^-J/\psi$ cross section at center-of-mass energies from 3.77 to 4.60~GeV,
	\href{Phys. Rev. Lett. 118, no.9, 092001 (2017)}{
	Phys. Rev. Lett. \textbf{118}, no.9, 092001 (2017)}.
	
	\bibitem{Liu:2020tqy}
	M.~Z.~Liu and L.~S.~Geng,
	Is $X(7200)$ the heavy anti-quark diquark symmetry partner of $ X(3872)$?,
	\href{Eur. Phys. J. C 81, no.2, 179 (2021)}{
	Eur. Phys. J. C \textbf{81}, no.2, 179 (2021)}.
	
	\bibitem{Wu:2023rrp}
	Q.~Wu, M.~Z.~Liu and L.~S.~Geng,
	Productions of \(X(3872)\), $Z_c(3900)$, $X_2(4013)$, and $Z_c(4020)$ in $B_{(s)}$ decays offer strong clues on their molecular nature,
	\href{Eur. Phys. J. C 84, no.2, 147 (2024)}{
	Eur. Phys. J. C \textbf{84}, no.2, 147 (2024)}.
	
	\bibitem{Liu:2024ziu}
	M.~Z.~Liu, X.~Z.~Ling and L.~S.~Geng,
	Productions of \(X(3872)/Z_c(3900)\) and \(X_2(4013)/Z_c(4020)\) in \(Y(4220)\) and \(Y(4360)\) decays,
	\href{Phys. Rev. D 110, no.5, 054035 (2024)}{
	Phys. Rev. D \textbf{110}, no.5, 054035 (2024)}.
	
	\bibitem{BESIII:2019tdo}
	M.~Ablikim \textit{et al.} [BESIII],
	Observation of $e^{+}e^{-}\rightarrow \pi^{+}\pi^{-}\psi(3770)$ and $D_{1}(2420)^{0}\bar{D}^{0} + c.c.$,
	\href{Phys. Rev. D 100, no.3, 032005 (2019)}{
	Phys. Rev. D \textbf{100}, no.3, 032005 (2019)}.
	
	
	\bibitem{Obraztsov:2016}
	V.~Obraztsov [OKA],
	High statistics measurement of the $K^+ \to \pi^0 e^+\nu$(Ke3) decay formfactors,
	\href{Nucl. Part. Phys. Proc. 273-275, 1330 (2016)}{
	Nucl. Part. Phys. Proc. \textbf{273-275}, 1330 (2016)}.
	
	
	\bibitem{Velghe:2016}
	B.~Velghe [NA62-RK and NA48/2],
	\(K^{\pm}\to \pi^{\pm}\gamma\gamma\) Studies at NA48/2 and NA62-RK Experiments at CERN,
	\href{Nucl. Part. Phys. Proc. 273-275, 2720 (2016)}{
	Nucl. Part. Phys. Proc. \textbf{273-275}, 2720 (2016)}.
	
	
	\bibitem{Quintans:2022utc}
	C.~Quintans [AMBER],
	The New AMBER Experiment at the CERN SPS,
	\href{Few Body Syst. 63, 72 (2022)}{
	Few Body Syst. \textbf{63}, 72 (2022)}.
	
	\bibitem{ATLAS:2012yve}
	G.~Aad \textit{et al.} [ATLAS],
	Observation of a new particle in the search for the Standard Model Higgs boson with the ATLAS detector at the LHC,
	\href{Phys. Lett. B 716, 1-29 (2012)}{
	Phys. Lett. B \textbf{716}, 1-29 (2012)}.
	
	\bibitem{CMS:2013btf}
	S.~Chatrchyan \textit{et al.} [CMS],
	Observation of a New Boson with Mass Near 125 GeV in $pp$ Collisions at $\sqrt{s}$ = 7 and 8 TeV,
	\href{JHEP 06, 081 (2013)}{
	JHEP \textbf{06}, 081 (2013)}.
	
	\bibitem{ALICE:2008ngc}
	K.~Aamodt \textit{et al.} [ALICE],
	The ALICE experiment at the CERN LHC,
	\href{JINST 3, S08002 (2008)}{
	JINST \textbf{3}, S08002 (2008)}.
	
	\bibitem{PHENIX:2001vgc}
	K.~Adcox \textit{et al.} [PHENIX],
	Centrality dependence of \(\pi^{+/-}\), \(K^{+/-}\), \(p\) and \(\bar{p}\) production from \(\sqrt{s_{NN}}=130\)~GeV Au+Au collisions at RHIC,
	\href{Phys. Rev. Lett. 88, 242301 (2002)}{
	Phys. Rev. Lett. \textbf{88}, 242301 (2002)}.
	
	\bibitem{FCC:2018vvp}
	A.~Abada \textit{et al.} [FCC],
	FCC-hh: The Hadron Collider: Future Circular Collider Conceptual Design Report Volume 3,
	\href{Eur. Phys. J. ST 228, no.4, 755-1107 (2019)}{
	Eur. Phys. J. ST \textbf{228}, no.4, 755-1107 (2019)}.
	
	
	\bibitem{Huang:2016ygf}
	Y.~Huang, J.~He, J.~J.~Xie and L.~S.~Geng,
	Production of the $\Lambda_c(2940)$ by kaon-induced reactions on a proton target,
	\href{Phys. Rev. D 99, no.1, 014045 (2019)}{
	Phys. Rev. D \textbf{99}, no.1, 014045 (2019)}.
	
	\bibitem{Huang:2020ptc}
	Y.~Huang, J.~X.~Lu, J.~J.~Xie and L.~S.~Geng,
	Strong decays of ${\bar{D}}^{*}K^{*}$ molecules and the newly observed $X_{0,1}$ states,
	\href{Eur. Phys. J. C 80, no.10, 973 (2020)}{
	Eur. Phys. J. C \textbf{80}, no.10, 973 (2020)}.
	
	
	\bibitem{Okubo:1975sc}
	S.~Okubo,
	SU(4), SU(8) Mass Formulas and Weak Interactions,
	\href{	Phys. Rev. D 11, 3261-3269 (1975)}{
	Phys. Rev. D \textbf{11}, 3261-3269 (1975)}.
	
	\bibitem{Lebiedowicz:2012}
	P.~Lebiedowicz and A.~Szczurek,
	$pp\rightarrow ppK^+K^-$ reaction at high energies,
	\href{Phys. Rev. D 85, 014026 (2012)}{
	Phys. Rev. D \textbf{85}, 014026 (2012)}. 
	
	
	\bibitem{Lebiedowicz:2013vya}
	P.~Lebiedowicz and A.~Szczurek,
	Exclusive $p p \to p p \pi^{0}$ reaction at high energies,
	\href{Phys. Rev. D 87, no.7, 074037 (2013)}{
	Phys. Rev. D \textbf{87}, no.7, 074037 (2013)}.
	
	
	\bibitem{Belle:2007qxm}
	G.~Pakhlova \textit{et al.} [Belle],
	Measurement of the near-threshold \(e^+e^-\to D\bar{D}\) cross section using initial-state radiation,
	\href{Phys. Rev. D 77, 011103 (2008)}{
	Phys. Rev. D \textbf{77}, 011103 (2008)}.
	
	
	\bibitem{BaBar:2006qlj}
	B.~Aubert \textit{et al.} [BaBar],
	Study of the Exclusive Initial-State Radiation Production of the \(D\bar{D}\) System,
	\href{Phys. Rev. D 76, 111105 (2007)}{
	Phys. Rev. D \textbf{76}, 111105 (2007)}.
	
	\bibitem{Guo:2016iej}
	X.~D.~Guo, D.~Y.~Chen, H.~W.~Ke, X.~Liu and X.~Q.~Li,
	Study on the rare decays of $Y(4630)$ induced by final state interactions,
	\href{Phys. Rev. D 93, no.5, 054009 (2016)}{
	Phys. Rev. D \textbf{93}, no.5, 054009 (2016)}.
	

	
	
	\bibitem{Garcilazo:2022edi}
	H.~Garcilazo and A.~Valcarce,
	Hidden-flavor pentaquarks,
	\href{Phys. Rev. D 106, no.11, 114012 (2022)}{
	Phys. Rev. D \textbf{106}, no.11, 114012 (2022)}.
	
	
	
	
	
	
	
	
	
\end{thebibliography}
\end{document}